\pdfoutput=1
\documentclass[11pt,draftcls,onecolumn]{IEEEtran}\newcommand{\matrixfontsize}{\fontsize{11}{11}\selectfont}

\usepackage[T1]{fontenc} 
\usepackage{amsmath} 
\usepackage{amssymb} 
\usepackage[mathscr]{eucal} 
\usepackage[final]{graphicx} 
\usepackage{color}
\usepackage{hyperref}
\IEEEeqnarraydefcolsep{15}{1.5\arraycolsep} 

\newcommand{\bsy}[1]{\boldsymbol{#1}} 
\newcommand{\diag}{\mbox{\rm diag}}
\newcommand{\gam}[1]{\mbox{$\gamma_{\!\ssst#1}$}} 
\newcommand{\order}{\mbox{\rm order}}
\newcommand{\ssst}{\scriptscriptstyle}
\newcommand{\suppint}{\mbox{\rm supp\_int}}	
\newcommand{\supprad}{\mbox{\rm supp\_rad}}

\newtheorem{cor}{Corollary} 
\newtheorem{exmp}{Example} 
\newtheorem{lem}{Lemma} 
\newtheorem{thm}{Theorem}

\newcommand{\case}{\paragraph*{Case}} 
\newcommand{\rem}{\paragraph*{Remarks}} 
\newcommand{\pf}{\IEEEproof} 
\newcommand{\qed}{\IEEEQED}  

\begin{document}
\date{December 9, 2009}
\title{Group Lifting Structures for Multirate Filter Banks,~II: Linear Phase Filter Banks}
\author{Christopher M.\ Brislawn
	\thanks{The author is with Los Alamos National Laboratory, Los Alamos, NM 87545--1663 USA (e-mail: brislawn@lanl.gov).   Los Alamos National Laboratory is operated by Los Alamos National Security LLC for the U.\ S.\ Department of Energy under contract DE-AC52-06NA25396.  This work was supported by the Los Alamos Laboratory-Directed Research \& Development Program.}\\
Los Alamos National Laboratory, MS B265, Los Alamos, NM 87545--1663 USA\\
(505) 665--1165 (office);\quad  (505) 665--5220 (FAX);\quad e-mail: {\tt brislawn@lanl.gov}
}
\maketitle
\vspace*{-0.25in}
\begin{center}
\textbf{\Large Final Revision, Reformatted for arXiv.org}
\end{center}
\vspace*{0.25in}

\begin{abstract}
The theory of group lifting structures  is applied to linear phase lifting factorizations for the two nontrivial classes of two-channel linear phase perfect reconstruction filter banks, the whole- and half-sample symmetric classes.  Group lifting structures defined for the reversible and irreversible classes of whole- and half-sample symmetric filter banks  are shown to satisfy the hypotheses of the uniqueness theorem for group lifting structures.  It follows that linear phase group lifting factorizations of whole- and half-sample symmetric filter banks are  therefore independent of the factorization methods used to construct them.  These results cover the  specification of whole-sample symmetric filter banks in the ISO/IEC JPEG~2000 image coding standard. 
\end{abstract}


\newpage\tableofcontents\clearpage


\section{Introduction}\label{sec:Intro}
This is the second paper  on a new approach to lifting factorization for two-channel  FIR perfect reconstruction filter banks.  The first paper~\cite{Bris09} introduced  \emph{group lifting structures} to parameterize  universes of  lifting factorizations for  given classes of filter banks.  While lifting factorizations are generally nonunique, it was proven in~\cite[Theorem~1]{Bris09} that, under suitable hypotheses, a filter bank has a \emph{unique} irreducible lifting factorization within a given group lifting structure.
The present paper applies~\cite[Theorem~1]{Bris09} to  group lifting structures for linear phase lifting factorizations of whole-sample symmetric (WS) and half-sample symmetric (HS) filter banks.  Existence of linear phase lifting factorizations for these classes  was proven in~\cite{BrisWohl06}, and group lifting structures  were introduced in~\cite[Section~IV]{Bris09} for both irreversible and reversible factorizations.  Using~\cite[Theorem~1]{Bris09}, we now prove uniqueness results for irreducible group lifting factorizations  of these  filter banks.  

For WS filter banks, like those in  the ISO/IEC JPEG~2000  standard~\cite[Annex~F]{ISO_15444_1}, \cite[Annex~G]{ISO_15444_2}, our results imply there is  only one way to factor them into  alternating upper and lower triangular lifting matrices with HS lifting filters.  Lifting HS filter banks from concentric, equal-length HS base filter banks with whole-sample antisymmetric (WA) lifting filters~\cite{BrisWohl06} is shown to be unique modulo one trivial degree of freedom (``unique modulo rescaling'').
The key to~\cite[Theorem~1]{Bris09} is verifying a  polyphase order-increasing hypothesis~\cite[Definition~10]{Bris09} by computing the order of lifted filter banks.  

The remainder of the Introduction reviews key notation and concepts from~\cite{Bris09}, adding some useful computational tools.  Section~\ref{sec:Sufficient} proves a technical result, Lemma~\ref{lem:SufficientConditions}, that gives sufficient conditions for  a group lifting structure to satisfy the order-increasing hypothesis.  Section~\ref{sec:WS} uses  Lemma~\ref{lem:SufficientConditions} to prove that the reversible and irreversible WS  group lifting structures satisfy the order-increasing property and generate unique group lifting factorizations.  Analogous results for HS group lifting structures are proven in Section~\ref{sec:HS}.  Section~\ref{sec: Conclusions} contains concluding remarks.

\subsection{Filter Banks}\label{sec:Intro:Banks}
Given an FIR filter,
\[ F(z) = \sum_{n=a}^b f(n)\,z^{-n}\,, \]
the {\em support interval\/} of $f$ is defined~\cite[Definition~1]{Bris09} to be the largest closed interval of integers for which $f(a),\, f(b)\neq 0$:
\begin{equation}\label{supp_int}
\suppint(F)\equiv \suppint(f)\equiv [a,b]\subset\mathbb{Z}\,.
\end{equation}
If {$\suppint(f)=[a,b]$} then the  \emph{order} of $F$ is defined to be
\begin{equation}\label{order}
\order(F) \equiv b-a\,. 
\end{equation}
Define the filter's \emph{support radius} to be the integer
\begin{equation}\label{supp_rad}
\supprad(f)\equiv \left\lfloor\frac{b-a+1}{2}\right\rfloor\,.
\end{equation}

The polyphase vector representation of an FIR filter,
\begin{IEEEeqnarray}{rCl}
\bsy{F}(z) & \equiv &
        \left[ \begin{IEEEeqnarraybox*}[\matrixfontsize][c]{,c,}
        F_0(z)\IEEEstrut[10pt][3pt]\\
        F_1(z)\IEEEstrut[10pt][3pt]
        \end{IEEEeqnarraybox*}\right]
=  \sum_{n=c}^{d}\bsy{f}(n)\,z^{-n},\label{poly_fltr}\\
\bsy{f}(n) & \equiv & 
        \left[ \begin{IEEEeqnarraybox*}[\matrixfontsize][c]{,c,}
        f_0(n)\IEEEstrut[10pt][3pt]\\
        f_1(n)\IEEEstrut[10pt][3pt]
        \end{IEEEeqnarraybox*}\right]
\quad\mbox{with $\bsy{f}(c),\,\bsy{f}(d)\neq\bsy{0}$\,,}\label{poly_fltr_impulse}
\end{IEEEeqnarray}
is defined in~\cite[Section~II-A]{Bris09} based on the analysis polyphase-with-advance representation of the scalar filter~\cite{BrisWohl06},
\[ F(z)=F_0(z^2) + zF_1(z^2) \,. \]
The \emph{polyphase} support interval of the filter with polyphase vector representation~(\ref{poly_fltr_impulse}) is defined in~\cite{Bris09} to be
\begin{equation}\label{poly_vector_supp_int}
\suppint(\bsy{f}) \equiv [c,d]  \,. 
\end{equation}

Define the \emph{join} of two closed intervals, denoted with a $\vee$,  to be the smallest closed interval containing their union:
\begin{equation}\label{interval_join}
[a,b]\vee [c,d] \equiv \left[\min(a,c),\,\max(b,d)\right]\,.
\end{equation}
It is straightforward to verify that the polyphase support interval~(\ref{poly_vector_supp_int}) is given by the join of the scalar support intervals for the two polyphase component filters:
\begin{equation}\label{poly_supp_int}
\suppint(\bsy{f}) = \suppint(f_0)\vee\suppint(f_1) \,.
\end{equation}

Similarly, consider an FIR filter bank with polyphase matrix
\begin{IEEEeqnarray}{rCl}
\mathbf{H}(z) & \equiv  & 
        \left[ \begin{IEEEeqnarraybox*}[\matrixfontsize][c]{,c,}
        \bsy{H}_0^T(z)\IEEEstrut[10pt][3pt]\\
        \bsy{H}_1^T(z)\IEEEstrut[11pt][3pt]
        \end{IEEEeqnarraybox*}\right]
= \sum_{n=c}^d \mathbf{h}(n)\,z^{-n}\,,\label{poly_matrix_def}\\
\mathbf{h}(n) & \equiv & 
        \left[ \begin{IEEEeqnarraybox*}[\matrixfontsize][c]{,c,}
        \bsy{h}_0^T(n)\IEEEstrut[10pt][3pt]\\
        \bsy{h}_1^T(n)\IEEEstrut[11pt][3pt]
        \end{IEEEeqnarraybox*}\right]
\quad\mbox{with $\mathbf{h}(c),\,\mathbf{h}(d)\neq \mathbf{0}$\,.}\label{poly_matrix_impulse_def}
\end{IEEEeqnarray}
The polyphase support interval of the filter bank~(\ref{poly_matrix_def},\,\ref{poly_matrix_impulse_def}) is defined  in~\cite[Section~II-A]{Bris09} to be
$\suppint(\mathbf{h}) \equiv [c,d] $.
Again, it is straightforward to verify that the polyphase support interval of the filter bank is  equal to the join of the polyphase support intervals of the two polyphase filter vectors:
\begin{equation}\label{FB_suppint}
\suppint(\mathbf{h}) = \suppint(\bsy{h}_0)\vee\suppint(\bsy{h}_1)\,.
\end{equation}

The following lemma describes the support intervals  that result from performing  elementary operations on FIR filters.  
\begin{lem}\label{lem:suppint}
If $F(z)$ and $G(z)$ are scalar FIR filters then
\begin{equation}\label{suppint_convolution}
\suppint(f*g) = \suppint(f) + \suppint(g)\,.
\end{equation}
(Note that~(\ref{suppint_convolution}) does \emph{not} extend to matrix-vector or matrix-matrix convolution; the lack of such a simple formula for matrix-matrix convolution is responsible for many of the technical difficulties in this paper.)
If $\mathbf{F}(z)$ is a matrix-, vector-, or scalar-valued FIR filter with \mbox{$\suppint(\mathbf{f})=[a,b]$} then
\begin{equation}\label{suppint_interpolation}
\suppint\left(\mathbf{F}(z^2)\right) = [2a,2b]\,.
\end{equation}
If $\mathbf{G}(z)$ has the same dimensions as $\mathbf{F}(z)$ and 
\mbox{$\suppint(\mathbf{g})\subset(a,b)=[a+1,\,b-1]$}, where $\subset$ denotes (non-proper) containment with possible equality, then
\begin{equation}\label{suppint_covering}
\suppint(\mathbf{f}+\mathbf{g}) = \suppint(\mathbf{f})\,.
\end{equation}
\end{lem}

\subsection{Lifting Factorizations}\label{sec:Intro:Lifting}
Part~I~\cite{Bris09} works with \emph{partially factored lifting cascades},
\begin{equation}\label{lift_from_B}
\mathbf{H}(z) = \diag(1/K,\,K)\,\mathbf{S}_{N-1}(z) \cdots\mathbf{S}_0(z)\, \mathbf{B}(z)\,,
\end{equation}
relative to a \emph{base} filter bank,  $\mathbf{B}(z)$.
Cascades that  are completely factored into lifting steps correspond to $\mathbf{B}(z)=\mathbf{I}$.
The \emph{update characteristic}, $m_n$, of  $\mathbf{S}_n(z)$~\cite[Definition~2]{Bris09} indicates whether a lifting matrix is upper triangular (i.e., a lowpass update) or lower triangular (a highpass update):
\begin{equation}\label{update_characteristic}
m_n = \left\{ 
\begin{array}{l}
0\quad\mbox{if $\mathbf{S}_n(z)$ is upper triangular,}\\
1\quad\mbox{if $\mathbf{S}_n(z)$ is lower triangular.}
\end{array}
\right.
\end{equation}
A lifting cascade is \emph{irreducible}~\cite[Definition~3]{Bris09} if 
the lifting steps strictly alternate between lower and upper triangular, implying that
$m_{n+1} = 1 - m_n$ for irreducible liftings.

Many of the arguments in this paper involve finite induction and are based on the recursive formulation of lifting:
\begin{IEEEeqnarray}{rCl}
\mathbf{H}(z) &=&  \diag(1/K,\,K)\,\mathbf{E}^{(N-1)}(z)\,,\IEEEnonumber\\
\mathbf{E}^{(n)}(z) &=& \mathbf{S}_n(z)\,\mathbf{E}^{(n-1)}(z)
		\mbox{\quad for $0\leq n< N$,}\label{recursive_cascade}\\
\mathbf{E}^{(-1)}(z)  &\equiv&  \mathbf{B}(z)\,.\IEEEnonumber
\end{IEEEeqnarray}
The  scalar filters corresponding to an intermediate partial product of lifting matrices  are  given by~\cite[eqn.~(9)]{BrisWohl06},
\begin{equation}\label{scalar_partial_products}
        \left[ \begin{IEEEeqnarraybox*}[\matrixfontsize][c]{c}
        E^{(n)}_0(z)\IEEEstrut[11pt][4pt]\\
        E^{(n)}_1(z)\IEEEstrut[11pt][3pt]
        \end{IEEEeqnarraybox*}\right]
= 
    \mathbf{E}^{(n)}(z^2)
        \left[ \begin{IEEEeqnarraybox*}[\matrixfontsize][c]{c}
        1\IEEEstrut[10pt][2pt]\\
        z\IEEEstrut[10pt][2pt]
        \end{IEEEeqnarraybox*}\right]
=
    \mathbf{S}_n(z^2)
        \left[ \begin{IEEEeqnarraybox*}[\matrixfontsize][c]{c}
        E^{(n-1)}_0(z)\IEEEstrut[11pt][4pt]\\
        E^{(n-1)}_1(z)\IEEEstrut[11pt][3pt]
        \end{IEEEeqnarraybox*}\right]\,,
\end{equation}
where ${E}^{(-1)}_i(z) \equiv  B_i(z)$ for $i=0,\,1.$
We make extensive use of the scalar version of this recursion, based on the update characteristic, $m_n$, of $\mathbf{S}_n(z)$.  The filter updated by $\mathbf{S}_n(z)$ is  $E^{(n)}_{m_n}(z)$, while $E^{(n)}_{1-m_n}(z)$ is \emph{not} modified by $\mathbf{S}_n(z)$.  Thus,~(\ref{scalar_partial_products}) corresponds to the scalar lifting formulas
\begin{IEEEeqnarray}{rCl}
E^{(n)}_{m_n}(z) &=& E^{(n-1)}_{m_n}(z) + S_n(z^2)\,E^{(n-1)}_{1-m_n}(z)\,,\label{En_scalar_lifting}\\
E^{(n)}_{1-m_n}(z) &=& E^{(n-1)}_{1-m_n}(z)\,.\label{En_scalar_lifting_complement}
\end{IEEEeqnarray}

\subsection{Group Lifting Structures}\label{sec:Intro:GLS}
A  \emph{group lifting structure}, $\mathfrak{S}$, is defined in~\cite[Definition~6]{Bris09} to be an ordered four-tuple, 
\begin{equation}\label{group_lifting_structure}
\mathfrak{S}\equiv(\mathscr{D},\,\mathscr{U},\,\mathscr{L},\,\mathfrak{B})\,. 
\end{equation}
$\mathscr{D}$ is an abelian group of diagonal gain scaling matrices,  \mbox{$\mathbf{D}_K \equiv \diag(1/K,\,K)$.}
$\mathscr{U}$ and $\mathscr{L}$ are abelian groups of upper and lower triangular lifting matrices, and $\mathfrak{B}$ is a set of base filter banks.  
The \emph{lifting cascade group},  $\mathscr{C}$,  is the nonabelian  matrix group generated by $\mathscr{U}$ and $\mathscr{L}$,
\[
\mathscr{C} \equiv \langle\mathscr{U\cup L}\rangle =
\left\{\rule[-1pt]{0pt}{12pt}\mathbf{S}_1\cdots\mathbf{S}_k \,:\, 
k\geq 1,\;\mathbf{S}_i\in\mathscr{U\cup L}\right\}\,.
\]
The universe  of all lifted filter banks generated by $\mathfrak{S}$ is 
\[ 
\mathscr{DC}\mathfrak{B} \equiv 
\left\{\rule[-1pt]{0pt}{12pt}\mathbf{DCB} : 
\mathbf{D}\in\mathscr{D},\;\mathbf{C}\in\mathscr{C}\,,\;\mathbf{B}\in\mathfrak{B}\right\}\,.
\]

The  matrix $\mathbf{D}_K\in\mathscr{D}$ acts on transfer matrices via the inner automorphism $\gam{K}$,
\begin{equation}\label{conjugation_operator}
\gam{K}\mathbf{A}(z) \equiv \mathbf{D}_K\,\mathbf{A}(z)\,\mathbf{D}_{K}^{-1} .
\end{equation}
A group $\mathscr{G}$ of transfer matrices is \emph{$\mathscr{D}$-invariant}~\cite[Definition~5]{Bris09} if $\gam{K}\mathscr{G} = \mathscr{G}$ for all $\mathbf{D}_K\in\mathscr{D}$, and a group lifting structure~(\ref{group_lifting_structure}) is $\mathscr{D}$-invariant if $\mathscr{U}$ and $\mathscr{L}$, and therefore $\mathscr{C}$, are $\mathscr{D}$-invariant.
$\mathscr{D}$-invariance  is easy to verify in practice using~\cite[Lemma~2]{Bris09}.

A lifting cascade~(\ref{lift_from_B}) is called \emph{strictly polyphase order-increasing} (or just \emph{order-increasing})~\cite[Definition~10]{Bris09} if the intermediate partial products~(\ref{recursive_cascade}) satisfy
\[ 
\order\left(\mathbf{E}^{(n)}\right) > \order\left(\mathbf{E}^{(n-1)}\right)\quad\mbox{for $0\leq n<N$.}
\]
A group lifting structure, $\mathfrak{S}$, is  called order-increasing if every irreducible cascade in 
\mbox{$\mathscr{C}\mathfrak{B}$} is order-increasing.

If $\mathfrak{S}$ is an order-increasing, $\mathscr{D}$-invariant group lifting structure,
the uniqueness theorem~\cite[Theorem~1]{Bris09} says that all irreducible factorizations of  $\mathbf{H}(z)$ in $\mathfrak{S}$ are ``equivalent modulo rescaling.''  Specifically, given two such factorizations,
\begin{eqnarray}
\mathbf{H}(z) & = & \mathbf{D}_K\,\mathbf{S}_{N-1}(z) \cdots\mathbf{S}_0(z)\, \mathbf{B}(z)  
                \label{unprimed_cascade} \\
        & = & \mathbf{D}_{K'}\,\mathbf{S}'_{N'-1}(z)\cdots \mathbf{S}'_0(z)\,\mathbf{B}'(z)\,,
                \label{primed_cascade}
\end{eqnarray}
the uniqueness theorem states that the number of lifting steps is the same ($N'=N$), with base filter banks related by
\begin{equation}\label{defn_alpha} 
\mathbf{B}'(z) = \mathbf{D}_{\alpha}\,\mathbf{B}(z)\quad\mbox{for constant\ }
\alpha= K/K',
\end{equation}
and lifting steps related by inner automorphisms,
\begin{equation}\label{almost_unique_factors} 
\mathbf{S}'_i(z) = \gamma_{\!\alpha}\mathbf{S}_i(z)\,,\quad\mbox{$i=0,\ldots,N-1$.}
\end{equation}
We express this by saying that irreducible lifting factorizations in $\mathfrak{S}$ are ``unique modulo rescaling''~\cite[Definition~11]{Bris09}.   

The theorem can be strengthened to yield \emph{unique} irreducible  factorizations by normalizing a degree of freedom in the base filter banks.  For instance, fixing a  lowpass DC response like  $B_0(1)=1$ for all $\mathbf{B}(z) \in \mathfrak{B}$ implies $\alpha=1$ by~(\ref{defn_alpha}).

Our principal tool for verifying the  order-increasing hypothesis of~\cite[Theorem~1]{Bris09} is developed in the next section.

\section{Sufficient Conditions for Order-Increasing Group Lifting Structures}
\label{sec:Sufficient}
Since  lifting only modifies one filter at a time, the  order-increasing property hinges on the growth of the support intervals of the filters $E^{(n)}_{m_n}(z)$ being updated in each lifting step.  A simplification  occurs in the WS and HS cases because, as we shall show, the \emph{polyphase} support interval of the updated polyphase filter vector, $\bsy{E}^{(n)}_{m_n}(z)$, always contains that of its complement, $\bsy{E}^{(n)}_{1-m_n}(z)$, making the polyphase order of $\mathbf{E}^{(n)}(z)$ equal to the {polyphase} order of $\bsy{E}^{(n)}_{m_n}(z)$.  This is convenient, but  other order-increasing group lifting structures may exist that do not share this property.
\begin{lem}\label{lem:SufficientConditions}
Let $\mathfrak{S}$ be a group lifting structure satisfying the following two polyphase vector conditions.

1)\quad For all $\mathbf{B}(z)\in\mathfrak{B}$, the polyphase support intervals~(\ref{poly_vector_supp_int}) for the base polyphase filter vectors are equal:
\begin{equation}\label{equal-order_base}
\suppint(\bsy{b}_0) = \suppint(\bsy{b}_1) \;.
\end{equation}

2)\quad For all irreducible lifting cascades in $\mathscr{C}\mathfrak{B}$, the  polyphase support intervals~(\ref{poly_vector_supp_int}) for the intermediate polyphase filter vectors satisfy the proper inclusions
\begin{equation}\label{support-covering}
\suppint\left(\bsy{e}^{(n)}_{1-m_n}\right) \varsubsetneq \suppint\left(\bsy{e}^{(n)}_{m_n}\right)
\quad\mbox{for $n\geq 0$.}
\end{equation}
It then follows that $\mathfrak{S}$ is strictly polyphase order-increasing.
\end{lem}
\pf
Let \mbox{$\mathbf{S}_{N-1}(z) \cdots\mathbf{S}_0(z)\, \mathbf{B}(z)\in\mathscr{C}\mathfrak{B}$} be irreducible.  
By~(\ref{FB_suppint}) and~(\ref{equal-order_base}) the base filter bank satisfies
\begin{IEEEeqnarray*}{rCl}
\suppint(\mathbf{b}) & = & \suppint(\bsy{b}_0)\vee\suppint(\bsy{b}_1)\\
&=& \suppint(\bsy{b}_0)\\
&=& \suppint(\bsy{b}_1)\,,
\end{IEEEeqnarray*}
which implies
\begin{equation}\label{EqualOrderBaseFilters}
\order(\mathbf{B}) = \order(\bsy{B}_0) = \order(\bsy{B}_1)\,.
\end{equation}
Similarly,~(\ref{FB_suppint}) and~(\ref{support-covering})  imply
\begin{IEEEeqnarray*}{rCl}
\suppint\left(\mathbf{e}^{(n)}\right) & = & \suppint\left(\bsy{e}^{(n)}_0\right) \vee \suppint\left(\bsy{e}^{(n)}_1\right)\\
&=& \suppint\left(\bsy{e}^{(n)}_{m_n}\right)\,,
\end{IEEEeqnarray*}
and therefore
\begin{equation}\label{IntermediateFilterBankOrder}
\order\left(\mathbf{E}^{(n)}\right) = \order\left(\bsy{E}^{(n)}_{m_n}\right)\,.
\end{equation}
Note that~(\ref{support-covering}) also implies
\begin{equation}\label{poly_order-increase}
\order\left(\bsy{E}^{(n)}_{m_n}\right) > \order\left(\bsy{E}^{(n)}_{1-m_n}\right)
\quad\mbox{for $n\geq 0$.}
\end{equation}

We now show that $\order\left(\mathbf{E}^{(n)}\right)$ is strictly increasing.

\case{$n=0.$}\quad
Since the filter \emph{not} being lifted is 
\mbox{$\bsy{E}^{(0)}_{1-m_0}(z)=\bsy{B}_{1-m_0}(z)$,}
by~(\ref{IntermediateFilterBankOrder}), (\ref{poly_order-increase}), and~(\ref{EqualOrderBaseFilters}) we get
\begin{IEEEeqnarray*}{l}
\order\left(\mathbf{E}^{(0)}\right)  =  \order\left(\bsy{E}^{(0)}_{m_0}\right)\\
\qquad >  \order\left(\bsy{E}^{(0)}_{1-m_0}\right)
=  \order\left(\bsy{B}_{1-m_0}\right)
= \order(\mathbf{B})\,.
\end{IEEEeqnarray*}
\case{$n>0.$}\quad
Since the cascade is irreducible by hypothesis,  the  filter \emph{not} being lifted in step~$n$ must have been lifted in step~$n-1$:
\[  \bsy{E}^{(n)}_{1-m_n}(z) = \bsy{E}^{(n-1)}_{m_{n-1}}(z)\,.  \]
Therefore,
\begin{IEEEeqnarray*}{l}
\order\left(\mathbf{E}^{(n)}\right)  =  \order\left(\bsy{E}^{(n)}_{m_n}\right)\\
\;\quad >  \order\left(\bsy{E}^{(n)}_{1-m_n}\right)
=  \order\left(\bsy{E}^{(n-1)}_{m_{n-1}}\right)
= \order\left(\mathbf{E}^{(n-1)}\right).
\end{IEEEeqnarray*}
\hfill\qed
\rem The polyphase representation of the  ``causal lazy wavelet'' filter bank~\cite[Example~7]{Bris09} does \emph{not} satisfy~(\ref{equal-order_base}).   A scalar analogue of~(\ref{support-covering}) became evident while computing linear phase lifting factorizations of both WS and HS filter banks, but the author found that~(\ref{equal-order_base}) was the less-than-obvious key to unifying the WS and HS cases.  
The base filters,
\begin{equation}\label{anticausal_lazy_wavelet}
B_0(z)=1\quad \mbox{and}\quad B_1(z)=z\,, 
\end{equation}
 for the (anticausal) lazy wavelet filter bank $\mathbf{B}(z)=\mathbf{I}$
 have equal lengths but are \emph{not} concentric like, e.g., the Haar filters, 
\begin{equation}\label{Haar_filters}
H_0(z)=(z+1)/2\quad\mbox{and}\quad H_1(z)=z-1\,. 
\end{equation}
An equal-length hypothesis for the base filters, though, is not enough by itself to imply the order-increasing property.  Fortunately,~(\ref{anticausal_lazy_wavelet}) and~(\ref{Haar_filters}) share a common feature that  implies the order-increasing property, namely, the fact that  in both cases the base filters have the same \emph{polyphase} support intervals.

\section{Whole-Sample Symmetric Filter Banks}\label{sec:WS}
We briefly review the definitions from~\cite[Section~IV-A]{Bris09}  of the group lifting structures for WS filter banks.  
For the irreversible WS group, $\mathscr{W}$, the  upper (respectively, lower) triangular lifting matrices, 
\mbox{$\mathscr{U} \equiv\upsilon(\mathscr{P}_0)$} (respectively, $\mathscr{L} \equiv\lambda(\mathscr{P}_1)$), are defined by   half-sample symmetric real Laurent polynomials:
\begin{IEEEeqnarray}{rCl}
\mathscr{P}_0 & \equiv& \left\{S(z)\in\mathbb{R}[z,z^{-1}] \,:\, S(z^{-1})=zS(z)\right\}\,,\label{WS_G0}\\
\mathscr{P}_1 & \equiv& \left\{S(z)\in\mathbb{R}[z,z^{-1}]  \,:\, S(z^{-1})=z^{-1}S(z)\right\}\,.\label{WS_G1}
\end{IEEEeqnarray}
The scaling matrices are $\mathscr{D} \equiv\left\{\mathbf{D}_K : K\neq 0\right\}$  and the base filter banks are trivial: $\mathfrak{B} \equiv\left\{\mathbf{I}\right\}$.  
The irreversible WS group lifting structure is $\mathfrak{S}_{\mathscr{W}}\equiv(\mathscr{D},\,\mathscr{U},\,\mathscr{L},\,\mathfrak{B})$, and its lifting cascade group  is $\mathscr{C_W\equiv\langle U\cup L\rangle}$.

In the reversible case there are no scaling matrices,  $\mathscr{D}_r \equiv\{\mathbf{I}\}$, the set of base filter banks is  still $\mathfrak{B}_r \equiv\left\{\mathbf{I}\right\}$, and the lifting filters~(\ref{WS_G0}), (\ref{WS_G1}) are restricted to those with dyadic rational coefficients.  The {reversible WS group} is  \emph{defined} by
\[ \mathscr{W}_r \equiv \mathscr{C}_{\mathscr{W}_r}\equiv\langle\mathscr{U}_r\cup \mathscr{L}_r\rangle \,, \]
and its group lifting structure is $\mathfrak{S}_{\mathscr{W}_r}\equiv(\mathscr{D}_r,\,\mathscr{U}_r,\,\mathscr{L}_r,\,\mathfrak{B}_r)$.

\subsection{Main Uniqueness Result}\label{sec:WS:Main}
Our main result on WS filter banks shows that $\mathfrak{S}_{\mathscr{W}}$ and  $\mathfrak{S}_{\mathscr{W}_r}$ produce unique irreducible group lifting factorizations.  
\begin{thm}[Uniqueness of WS group lifting factorizations]\label{thm:WS_Uniqueness}
Let $\mathfrak{S}_{\mathscr{W}}$ and $\mathfrak{S}_{\mathscr{W}_r}$ be the group lifting structures defined in~\cite[Section~IV-A]{Bris09}.
Every filter bank in $\mathscr{W}$ has a unique irreducible lifting factorization in $\mathfrak{S}_{\mathscr{W}}$ and every filter bank in $\mathscr{W}_r$ has a unique irreducible lifting factorization in $\mathfrak{S}_{\mathscr{W}_r}$.
\end{thm}

\pf
Existence of irreducible WS group lifting factorizations for these groups was covered in~\cite[Section~IV-A]{Bris09}.

Uniqueness:  As discussed above, $\mathfrak{B} =\left\{\mathbf{I}\right\}$ satisfies hypothesis~(\ref{equal-order_base})  in Lemma~\ref{lem:SufficientConditions} since
\[ \suppint(\bsy{b}_i) = \{0\}\,,\quad i=0,\,1. \]
Lemma~\ref{lem:WS_support_covering} (Section~\ref{sec:WS:UniqueFact} below) shows that irreducible lifting cascades in  $\mathscr{C_W}\mathfrak{B}$ satisfy~(\ref{support-covering}), so Lemma~\ref{lem:SufficientConditions} implies that  $\mathfrak{S}_{\mathscr{W}}$ is  order-increasing.
$\mathscr{D}$-invariance of  $\mathscr{U}$ and $\mathscr{L}$ follows from~\cite[Lemma~2]{Bris09}, so~\cite[Theorem~1]{Bris09} applies to $\mathfrak{S}_{\mathscr{W}}$.  
Since $\mathfrak{B}=\{\mathbf{I}\}$, it follows that the degree of freedom, $\alpha$, in the conclusion of~\cite[Theorem~1]{Bris09}  must be unity:  $\alpha=1$.  This means that irreducible group lifting factorizations in  $\mathfrak{S}_{\mathscr{W}}$ are unique.

In the reversible case we have 
$\mathscr{C}_{\mathscr{W}_r}<\mathscr{C}_{\mathscr{W}}$ and $\mathfrak{B}_r=\mathfrak{B}$.  The \emph{reversible} WS group lifting factorizations are thus a subset of the  \emph{irreversible}  factorizations,
\mbox{$\mathscr{C}_{\mathscr{W}_r}\mathfrak{B}_r \subset \mathscr{D}\mathscr{C}_\mathscr{W}\mathfrak{B}$,}  
so uniqueness of irreducible factorizations in $\mathfrak{S}_{\mathscr{W}_r}$ follows from uniqueness of irreducible factorizations in $\mathfrak{S}_\mathscr{W}$.  
\hfill\qed

\rem
While  existence of cascade factorizations for arbitrary WS filter banks was asserted for the  cascade structure in~\cite{NguyenVaidyana:89:Two-channel-perfect-reconstruction-structures},
Theorem~\ref{thm:WS_Uniqueness} provides a more complete theory of both existence \emph{and} uniqueness for factorizations of WS filter banks.

\subsubsection{Applications}\label{sec:WS:Main:Applications}
Theorem~\ref{thm:WS_Uniqueness}  implies that compliant lifting factorizations of
user-defined WS filter banks in JPEG~2000  Part~2, which  are specified in~\cite[Annex~G]{ISO_15444_2} as irreducible  WS group lifting factorizations, are \emph{unique}.  
\begin{cor}\label{cor:J2K_Pt2_Annex_G_uniqueness}
A delay-minimized unimodular WS filter bank can be specified in JPEG~2000 Part~2 Annex~G syntax in one and only one way.
\end{cor}

\begin{exmp}\label{exmp:WS_least_dissimilar_length}
In their book on JPEG~2000~\cite[Section~6.4.4]{TaubMarc02}, Taubman and Marcellin note the relative simplicity of implementing  lifting factorizations that only involve HS lifting filters of length two.  (It is clear from~\cite[Figure~6.13]{TaubMarc02} that this means \emph{first-order}  filters; i.e., symmetric filters with two consecutive terms, such as $c(1+z^{-1})$.)   They then  claim~\cite[page~294]{TaubMarc02}:
\begin{quote}
All two channel FIR subband transforms having odd length, symmetric filters with least dissimilar lengths (filter lengths differ by 2) may be factored into lifting steps of this form \emph{[i.e., first-order HS lifting filters].}
\end{quote}
As an application of Theorem~\ref{thm:WS_Uniqueness}, we can construct  counterexamples to this claim.
Consider the following lifting cascade:
\begin{equation}\label{non-first-order_lifting}
\mathbf{H}(z) =
        \left[ \begin{IEEEeqnarraybox*}[\matrixfontsize][c]{,c15c,}
                1  & 1+z^{-1} \\
                0    & 1%
        \end{IEEEeqnarraybox*}\right]
        \left[ \begin{IEEEeqnarraybox*}[\matrixfontsize][c]{,c15c,}
                1  & 0 \\
                z^2+z+1+z^{-1}  & 1
        \end{IEEEeqnarraybox*}\right]\,.
\end{equation}
By~(\ref{scalar_partial_products}) the corresponding WS scalar filter bank is
\begin{equation}\label{scalar_filters}
        \begin{array}{rcl}
        H_0(z) &=& z^4+2z^2+z+3+z^{-1}+2z^{-2}+z^{-4}\\
        H_1(z) &=& z^4+z^2+z+1+z^{-2}\,,
        \end{array}
\end{equation}
which has filters of least dissimilar lengths.  Theorem~\ref{thm:WS_Uniqueness} implies that~(\ref{non-first-order_lifting}) is its unique irreducible WS group lifting factorization, so it cannot have another such factorization using first-order HS filters.
Generalizing this example, it follows from Lemma~\ref{lem:WS_support_formula} (below) that an irreducible WS group lifting cascade corresponds to  a WS filter bank with filters of least dissimilar lengths if and only if the \emph{final} HS lifting filter is  first-order.
\end{exmp}
\smallskip

The lemmas needed to verify hypothesis~(\ref{support-covering}) of Lemma~\ref{lem:SufficientConditions} occupy the remainder of Section~\ref{sec:WS}.

\subsection{Polyphase Support Intervals}\label{sec:WS:Support}
The relation between a filter's impulse response, $e(k)$, and its polyphase components, $e_j(k)$,  is given in~\cite[eqn.~(10)]{BrisWohl06}:
\begin{equation}\label{poly_impulse_response}
e_j(k) = e(2k-j)\,,\quad j=0,\,1,\quad k\in\mathbb{Z}\,.
\end{equation}
In general, the support intervals~(\ref{supp_int}) for the individual polyphase components are not completely determined by $\suppint(e)$ (see the proof of the following lemma).  Nonetheless,  the \emph{polyphase} support interval, $\suppint(\bsy{e})$, \emph{is} completely determined by $\suppint(e)$.  The following derivations do not depend on linear phase properties of the filters.

\begin{lem}\label{lem:0_poly_suppint}
Suppose $E(z)$ is an odd-length FIR filter whose support interval~(\ref{supp_int}) is centered at 0,
\begin{equation}\label{0_suppint}
\suppint(e) = [-r,\,r]\,,
\end{equation}
where $r\equiv\supprad(e)\geq 0$ is  the support radius~(\ref{supp_rad}).  Then the \emph{polyphase} support interval~(\ref{poly_vector_supp_int})  of the filter is:
\[ 
\suppint(\bsy{e}) = \left\{
\begin{array}{ll}
[-r/2,\, r/2]  & \mbox{if $r$ is even,} \\
\left[(-r+1)/2,\, (r+1)/2\right]  & \mbox{if $r$ is odd.}
\end{array}
\right.
\]

\end{lem}
\pf   There are two cases.
\case{\it $r$ Even.}\quad
By~(\ref{0_suppint}) and~(\ref{poly_impulse_response}) the even polyphase component is 
$e_0(k)=e(2k)$ for $-r\leq 2k\leq r$,
and both $e(-r)$ and $e(r)$ are nonzero by hypothesis~(\ref{0_suppint}), so 
\begin{equation}\label{0_poly_suppint_even_0}
\suppint(e_0) = [-r/2,\,r/2]\,.
\end{equation}

The odd polyphase component is 
$e_1(k)=e(2k-1)$ for \mbox{$-r+1\leq 2k-1\leq r-1$,}
 but we don't know that $e$ is nonzero at the endpoints since they lie in the interior of $\suppint(e)$.  For instance, Strang and Nguyen~\cite[Section~6.5]{StrNgu96} exhibit a family of dyadic WS wavelet filter banks, credited to Sweldens, whose lowpass filters have \mbox{$h(\pm(r-1))=0$}, and~(\ref{scalar_filters}) provides another example.  
Thus, all we can conclude is that
\begin{equation}\label{0_poly_suppint_even_1}
\suppint(e_1) \subset [(-r+2)/2,\,r/2]\,.
\end{equation}
Combining~(\ref{0_poly_suppint_even_0}) and~(\ref{0_poly_suppint_even_1}) using~(\ref{poly_supp_int}) yields
\begin{IEEEeqnarray}{rCl}
\suppint(\bsy{e}) &=& \suppint(e_0)\vee\suppint(e_1)\IEEEnonumber \\
&=&[-r/2,\,r/2]\,.\label{0_poly_suppint_even}
\end{IEEEeqnarray}
\case{\it  $r$ Odd.}\quad
By~(\ref{0_suppint}) and~(\ref{poly_impulse_response}) the even polyphase component is 
$e_0(k)=e(2k)$ for \mbox{$-r+1\leq 2k\leq r-1$,}
and $e$ may vanish at either endpoint  since they both lie in the interior of $\suppint(e)$, so 
\begin{equation}\label{0_poly_suppint_odd_0}
\suppint(e_0) \subset [(-r+1)/2,\,(r-1)/2]\,.
\end{equation}

The odd polyphase component is 
$e_1(k)=e(2k-1)$ for \mbox{$-r\leq 2k-1\leq r$,}
and both $e(-r)$ and $e(r)$ are nonzero so
\begin{equation}\label{0_poly_suppint_odd_1}
\suppint(e_1) = [(-r+1)/2,\,(r+1)/2]\,.
\end{equation}
Combining~(\ref{0_poly_suppint_odd_0}) and~(\ref{0_poly_suppint_odd_1}) using~(\ref{poly_supp_int}) yields
\begin{IEEEeqnarray}{rCl}
\suppint(\bsy{e})& =& \suppint(e_0)\vee\suppint(e_1) \IEEEnonumber \\
&=&  [(-r+1)/2,\,(r+1)/2]\,.\label{0_poly_suppint_odd}
\end{IEEEeqnarray}
\hfill\qed
%
%
\begin{lem}\label{lem:-1_poly_suppint}
Suppose $E(z)$ is an odd-length FIR filter whose support interval is centered at $-1$:
\begin{equation}\label{-1_suppint}
\suppint(e) = [-r-1,\,r-1]\,,\quad  r\equiv\supprad(e)\geq 0\,.
\end{equation}
Then the polyphase support interval of the filter is:
\[ 
\suppint(\bsy{e}) = \left\{
\begin{array}{ll}
[-r/2,\, r/2]  & \mbox{if $r$ is even,} \\
\left[(-r-1)/2,\, (r-1)/2\right]  & \mbox{if $r$ is odd.}
\end{array}
\right.
\]
\end{lem}
\pf There are two cases.
\case{\it  $r$ Even.}\quad
The even polyphase component is 
$e_0(k)=e(2k)$ for \mbox{$-r\leq 2k\leq r-2$,} 
and $e$ may vanish at either endpoint  since both lie in the interior of $\suppint(e)$, so 
\begin{equation}\label{-1_poly_suppint_even_0}
\suppint(e_0) \subset [-r/2,\,(r-2)/2]\,.
\end{equation}

The odd polyphase component is 
$e_1(k)=e(2k-1)$ for \mbox{$-r-1\leq 2k-1\leq r-1$,}
 and \mbox{$e_1(k)\neq 0$} at both endpoints so
\begin{equation}\label{-1_poly_suppint_even_1}
\suppint(e_1) = [-r/2,\,r/2]\,.
\end{equation}
Combining~(\ref{-1_poly_suppint_even_0}) and~(\ref{-1_poly_suppint_even_1}) using~(\ref{poly_supp_int}) yields
\begin{equation}\label{-1_poly_suppint_even}
\suppint(\bsy{e}) =   [-r/2,\,r/2]\,.
\end{equation}
\case  \emph{$r$ Odd.}\quad
The even polyphase component is 
$e_0(k)=e(2k)$ for \mbox{$-r-1\leq 2k\leq r-1$,}
and $e$ is nonzero at both endpoints so 
\begin{equation}\label{-1_poly_suppint_odd_0}
\suppint(e_0) = [(-r-1)/2,\,(r-1)/2]\,.
\end{equation}

The odd polyphase component is 
$e_1(k)=e(2k-1)$ for \mbox{$-r\leq 2k-1\leq r-2$,} 
and $e$ may vanish at either endpoint  since they both lie in the interior of $\suppint(e)$ so
\begin{equation}\label{-1_poly_suppint_odd_1}
\suppint(e_1) \subset [(-r+1)/2,\,(r-1)/2]\,.
\end{equation}
Combining~(\ref{-1_poly_suppint_odd_0}) and~(\ref{-1_poly_suppint_odd_1}) using~(\ref{poly_supp_int}) yields
\begin{equation}\label{-1_poly_suppint_odd}
\suppint(\bsy{e}) =   [(-r-1)/2,\,(r-1)/2]\,.
\end{equation}
\hfill\qed

\subsection{The WS Support Interval Formula}\label{sec:WS:Formula}
We now calculate the support intervals of the intermediate filters, $E^{(n)}_i(z)$, for irreducible lifting cascades in  $\mathfrak{S}_{\mathscr{W}}$.  Although stated for the case of interest---WS group lifting cascades---the next lemma only uses the  support intervals for the filters and does not actually depend on the symmetry of either the  intermediate WS filters or  the HS lifting filters.

\begin{lem}\label{lem:WS_support_formula}
Let \mbox{$\mathbf{S}_{N-1}(z) \cdots\mathbf{S}_0(z)\in\mathscr{C}_{\mathscr{W}}$} be an irreducible cascade with intermediate  scalar filters $E^{(n)}_i(z),$ $i=0,\,1.$  Let $r^{(n)}_i$ be the support radius of  $e^{(n)}_i,$ 
and let \mbox{$t^{(n)}\geq 1$} be the support radius of the HS lifting filter $S_n(z)$.
Then $\suppint\left(e^{(n)}_i\right)$ is centered at $-i$,
\[ \suppint\left(e^{(n)}_i\right) = \left[-r^{(n)}_i-i,\,r^{(n)}_i-i\right]\,,\quad i=0,\,1, \]
where
\begin{equation}\label{WS_support_radius}
r^{(n)}_{m_n} = r^{(n)}_{1-m_n} + 2t^{(n)} -1\quad\mbox{for $n\geq 0$,}
\end{equation}
\begin{equation}\label{WS_support_radius_nonlift}
r^{(n)}_{1-m_n} = r^{(n-1)}_{m_{n}} + 2t^{(n-1)} -1\quad\mbox{for $n\geq 1$,} 
\end{equation}
with $r^{(0)}_{1-m_0} = r^{(-1)}_{1-m_0}=0$.
\end{lem}
\pf  Induction on $n$.
\case{$n=0,\; m_0=0.$}\quad
By~(\ref{WS_G0}) the lifting filter is centered at 1/2:
$\suppint(s_0)=[-t^{(0)}+1,\,t^{(0)}].$
By~(\ref{En_scalar_lifting}) the lifted intermediate filter is 
\begin{equation}\label{lifted_e00}
E^{(0)}_0(z) = 1 + z\,S_{0}(z^2)
\end{equation}
and its complement is
$E^{(0)}_1(z) = B_1(z) = z$, which implies \mbox{$r^{(0)}_{1} = 0$.}
Since $1=z^0$ we have $\suppint(z^0)=\{0\}$.  By~(\ref{suppint_convolution}) and~(\ref{suppint_interpolation})
\begin{eqnarray*}
\suppint\left(z\,S_{0}(z^2)\right) &=& \{-1\} + [2(-t^{(0)}+1),\,2t^{(0)}] \\
&=& [-2t^{(0)}+1,\,2t^{(0)}-1]\,. 
\end{eqnarray*}
Thus, \mbox{$t^{(0)}\geq 1$} implies 
\[ \suppint(z^0) \subset (-2t^{(0)}+1,\,2t^{(0)}-1)\,, \]
which justifies applying formula~(\ref{suppint_covering}) from Lemma~\ref{lem:suppint} to~(\ref{lifted_e00}):
\begin{eqnarray*}
\suppint\left(e^{(0)}_0\right) &=& \suppint\left(z\,S_{0}(z^2)\right) \\
&=&  [-2t^{(0)}+1,\,2t^{(0)}-1]\,.
\end{eqnarray*}
We conclude that $r^{(0)}_0 = 2t^{(0)}-1$, which verifies~(\ref{WS_support_radius}) since $r^{(0)}_{1} = 0$.
\smallskip

\case{$n=0,\; m_0=1.$}\quad
By~(\ref{WS_G1}) the lifting filter is centered at $-1/2$:
$\suppint(s_0)=[-t^{(0)},\,t^{(0)}-1].$
By~(\ref{En_scalar_lifting}) the lifted intermediate filter is 
\begin{equation}\label{lifted_e01}
E^{(0)}_1(z) = z + S_{0}(z^2)
\end{equation}
and its complement is
$E^{(0)}_0(z) = B_0(z) = 1$, which implies \mbox{$r^{(0)}_{0} = 0$.}
Since  $\suppint(z)=\{-1\}$ and 
\[ \suppint\left(S_{0}(z^2)\right) =  [-2t^{(0)},\,2(t^{(0)}-1)]\,, \]
\mbox{$t^{(0)}\geq 1$} implies
\[ \suppint(z) \subset (-2t^{(0)},\,2(t^{(0)}-1))\,, \]
which justifies 
applying formula~(\ref{suppint_covering})  to~(\ref{lifted_e01}):
\begin{eqnarray*}
\suppint\left(e^{(0)}_1\right) &=& \suppint\left(S_{0}(z^2)\right) \\
&=&  [-(2t^{(0)}-1)-1,\,(2t^{(0)}-1)-1]\,.
\end{eqnarray*}
We conclude that $r^{(0)}_1 = 2t^{(0)}-1$, which verifies~(\ref{WS_support_radius}) since $r^{(0)}_{0} = 0$.
\smallskip

\case{$n>0,\; m_n=0.$}\quad
The case $m_n=0$ means that 
\[ E^{(n)}_1(z)=E^{(n-1)}_1(z)\,, \]
and the induction hypothesis implies that  $\suppint\left(e^{(n)}_{1}\right)$ is centered at $-1$.
The cascade is assumed irreducible so $E^{(n-1)}_1(z)$ must have been lifted in step $n-1$, and the induction hypothesis therefore implies 
\begin{equation}\label{rn1_induction}
r^{(n)}_1 = r^{(n-1)}_1 = r^{(n-1)}_0 + 2t^{(n-1)} - 1\,, 
\end{equation}
which verifies~(\ref{WS_support_radius_nonlift}).  
By~(\ref{En_scalar_lifting}) the lifted intermediate filter is 
\begin{equation}\label{lifted_en0}
E^{(n)}_0(z) = E^{(n-1)}_0(z) + S_{n}(z^2)\,E^{(n-1)}_1(z)\,.
\end{equation}
$S_n(z)$ is centered at 1/2 so we have
\[ \suppint(s_n)=[-t^{(n)}+1,\,t^{(n)}]\,, \]
and~(\ref{suppint_convolution}) and~(\ref{suppint_interpolation}) give
\begin{IEEEeqnarray}{l}
\suppint\left(S_{n}(z^2)\,E^{(n-1)}_1(z)\right) \IEEEnonumber\\
\qquad = [2(-t^{(n)}+1),\,2t^{(n)}] + [-r^{(n-1)}_1-1,\,r^{(n-1)}_1-1] \IEEEnonumber\\
\qquad =  \left[-r^{(n-1)}_1 - 2t^{(n)} + 1,\; r^{(n-1)}_1 + 2t^{(n)} - 1\right]\,. \label{suppint_lowpass_lift}
\end{IEEEeqnarray}
Substitution of~(\ref{rn1_induction}) into~(\ref{suppint_lowpass_lift}) gives
\begin{IEEEeqnarray*}{l}
\suppint\left(S_{n}(z^2)\,E^{(n-1)}_1(z)\right)  \\
\qquad \qquad = 
\left[- r^{(n-1)}_0 -  2(t^{(n-1)} + t^{(n)} - 1)\,,\right.\\
\qquad \qquad \qquad \left. 
r^{(n-1)}_0 + 2(t^{(n-1)} + t^{(n)} - 1)\right]\,.
\end{IEEEeqnarray*}
The induction hypothesis says that $\suppint\left(e^{(n-1)}_0\right)$ is centered at 0,
with $t^{(n-1)}\geq 1$ and $t^{(n)}\geq 1$, so
\begin{IEEEeqnarray*}{rCl}
\suppint\left(e^{(n-1)}_0\right)  &=& \left[- r^{(n-1)}_0,\,  r^{(n-1)}_0\right] \\
&\subset&
\left(- r^{(n-1)}_0 -  2(t^{(n-1)} + t^{(n)} - 1)\,,\right.\\
&&\quad\left. r^{(n-1)}_0 + 2(t^{(n-1)} + t^{(n)} - 1)\right)\,.
\end{IEEEeqnarray*}
This justifies applying~(\ref{suppint_covering}) to~(\ref{lifted_en0}):
\[  \suppint\left(e^{(n)}_0\right) = \suppint\left(S_{n}(z^2)\,E^{(n-1)}_1(z)\right)\,. \]
Formula~(\ref{suppint_lowpass_lift}) implies that $\suppint\left(e^{(n)}_0\right)$ is centered at 0 and,
since $r^{(n)}_1=r^{(n-1)}_1$, it  verifies~(\ref{WS_support_radius}):
\[ r^{(n)}_0 = r^{(n)}_1+ 2t^{(n)} - 1\,. \]

\case{$n>0,\; m_n=1.$}\quad
The case $m_n=1$ means that 
\[ E^{(n)}_0(z)=E^{(n-1)}_0(z)\,, \]
and the induction hypothesis implies that  $\suppint\left(e^{(n)}_{0}\right)$ is centered at $0$.
The cascade is assumed irreducible so $E^{(n-1)}_0(z)$ must have been lifted in step $n-1$, and the induction hypothesis therefore implies 
\begin{equation}\label{rn0_induction}
r^{(n)}_0 = r^{(n-1)}_0 = r^{(n-1)}_1 + 2t^{(n-1)} - 1\,,
\end{equation}
which verifies~(\ref{WS_support_radius_nonlift}).  
By~(\ref{En_scalar_lifting}) the lifted intermediate filter is 
\begin{equation}\label{lifted_en1}
E^{(n)}_1(z) = E^{(n-1)}_1(z) + S_{n}(z^2)\,E^{(n-1)}_0(z)\,.
\end{equation}
$S_n(z)$ is centered at $-1/2$ so we have 
\[ \suppint(s_n)=[-t^{(n)},\,t^{(n)}-1]\,, \]
and~(\ref{suppint_convolution}) and~(\ref{suppint_interpolation}) give
\begin{IEEEeqnarray}{l}
\suppint\left(S_{n}(z^2)\,E^{(n-1)}_0(z)\right) \IEEEnonumber\\
\qquad = [-2t^{(n)},\,2(t^{(n)}-1)] + [-r^{(n-1)}_0,\,r^{(n-1)}_0] \IEEEnonumber\\
\qquad=  \left[-r^{(n-1)}_0 - 2t^{(n)},\; r^{(n-1)}_0 + 2t^{(n)} - 2\right]\,. \label{suppint_hipass_lift}
\end{IEEEeqnarray}
Substitution of~(\ref{rn0_induction}) into~(\ref{suppint_hipass_lift}) gives
\begin{IEEEeqnarray*}{l}
\suppint\left(S_{n}(z^2)\,E^{(n-1)}_0(z)\right)  \\
\qquad\qquad= \left[- r^{(n-1)}_1 -  2(t^{(n-1)} + t^{(n)}) + 1\,,\right.\\
\qquad\qquad \qquad\left. r^{(n-1)}_1 + 2(t^{(n-1)} + t^{(n)} )- 3\right]\,.
\end{IEEEeqnarray*}
The induction hypothesis says that $\suppint\left(e^{(n-1)}_1\right)$ is centered at $-1$,
with $t^{(n-1)}\geq 1$ and $t^{(n)}\geq 1$, so
\begin{IEEEeqnarray*}{rCl}
\suppint\left(e^{(n-1)}_1\right)  &=& \left[- r^{(n-1)}_1-1,\,  r^{(n-1)}_1-1\right] \\
&\subset&
\left(- r^{(n-1)}_1 -  2(t^{(n-1)} + t^{(n)}) + 1\,,\right.\\
&&\quad\left. r^{(n-1)}_1 + 2(t^{(n-1)} + t^{(n)}) - 3\right)\,,
\end{IEEEeqnarray*}
which justifies applying~(\ref{suppint_covering}) to~(\ref{lifted_en1}):
\[  \suppint\left(e^{(n)}_1\right) = \suppint\left(S_{n}(z^2)\,E^{(n-1)}_0(z)\right)\,. \]
Formula~(\ref{suppint_hipass_lift}) implies that $\suppint\left(e^{(n)}_1\right)$ is centered at $-1$ and,
since $r^{(n)}_0=r^{(n-1)}_0$,  it verifies~(\ref{WS_support_radius}):
\[ r^{(n)}_1 = r^{(n)}_0+ 2t^{(n)} - 1\,. \]
\hfill\qed

\subsection{The Support-Covering Property}\label{sec:WS:UniqueFact}
Now we can prove that $\mathfrak{S}_{\mathscr{W}}$ satisfies the polyphase support-covering hypothesis~(\ref{support-covering}) in Lemma~\ref{lem:SufficientConditions}.
\begin{lem}\label{lem:WS_support_covering}
Let \mbox{$\mathbf{S}_{N-1}(z) \cdots\mathbf{S}_0(z)\in\mathscr{C}_{\mathscr{W}}$} be irreducible, with intermediate  filters $E^{(n)}_i(z),$ $i=0,\,1.$  Then
\begin{equation}\label{WS_support-covering}
\suppint\left(\bsy{e}^{(n)}_{1-m_n}\right) \varsubsetneq \suppint\left(\bsy{e}^{(n)}_{m_n}\right)
\quad\mbox{for $n\geq 0$.}
\end{equation}
\end{lem}
\pf  Let $r^{(n)}_i$ be the support radius of $e^{(n)}_i$, $i=0,\,1$.  By Lemma~\ref{lem:WS_support_formula} it follows that
\mbox{$r^{(n)}_{m_n} + r^{(n)}_{1-m_n}$} is odd, so $r^{(n)}_{0}$ and $r^{(n)}_{1}$ always have \emph{opposite} parities.

\case{\it $r^{(n)}_0$ even, $r^{(n)}_1$ odd.}\quad
By Lemma~\ref{lem:0_poly_suppint} and Lemma~\ref{lem:-1_poly_suppint} the polyphase support intervals are:
\begin{IEEEeqnarray}{rCl}
\suppint\left(\bsy{e}^{(n)}_0\right) &=& \left[-r^{(n)}_0/2,\,r^{(n)}_0/2\right]\,, \label{even_odd_0}\\
\suppint\left(\bsy{e}^{(n)}_1\right) &=& \left[-\left(r^{(n)}_1+1\right)/2,\,\left(r^{(n)}_1-1\right)/2\right].\quad \label{even_odd_1}
\end{IEEEeqnarray}
If $m_n=0$ then applying Lemma~\ref{lem:WS_support_formula} to~(\ref{even_odd_0}) gives
\begin{IEEEeqnarray*}{l}
\suppint\left(\bsy{e}^{(n)}_0\right) \\
\qquad= \left[-\left(r^{(n)}_1+2t^{(n)}-1\right)/2,\,\left(r^{(n)}_1+2t^{(n)}-1\right)/2\right]\,. 
\end{IEEEeqnarray*}
Since $t^{(n)}\geq 1$, comparison with~(\ref{even_odd_1}) yields
\begin{IEEEeqnarray*}{rCl}
-\left(r^{(n)}_1+2t^{(n)}-1\right)/2 & \leq & -\left(r^{(n)}_1+1\right)/2\quad \mbox{and}\\
\left(r^{(n)}_1+2t^{(n)}-1\right)/2 & > & \left(r^{(n)}_1-1\right)/2\,,
\end{IEEEeqnarray*}
which imply $\suppint\left(\bsy{e}^{(n)}_1\right) \varsubsetneq \suppint\left(\bsy{e}^{(n)}_0\right)$.

If $m_n=1$ then applying Lemma~\ref{lem:WS_support_formula} to~(\ref{even_odd_1}) gives
\begin{IEEEeqnarray*}{l}
\suppint\left(\bsy{e}^{(n)}_1\right) \\
\qquad= \left[-\left(r^{(n)}_0+2t^{(n)}\right)/2,\,\left(r^{(n)}_0+2t^{(n)}-2\right)/2\right]\,.
\end{IEEEeqnarray*}
Since $t^{(n)}\geq 1$, comparison with~(\ref{even_odd_0}) yields
\begin{IEEEeqnarray*}{rCl}
-\left(r^{(n)}_0+2t^{(n)}\right)/2 & < & -r^{(n)}_0/2\quad \mbox{and}\\
\left(r^{(n)}_0+2t^{(n)}-2\right)/2 & \geq & r^{(n)}_0/2\,,
\end{IEEEeqnarray*}
which imply $\suppint\left(\bsy{e}^{(n)}_0\right) \varsubsetneq \suppint\left(\bsy{e}^{(n)}_1\right)$.

\case{\it $r^{(n)}_0$ odd, $r^{(n)}_1$ even.}\quad
By Lemma~\ref{lem:0_poly_suppint} and Lemma~\ref{lem:-1_poly_suppint} the polyphase support intervals are:
\begin{IEEEeqnarray}{rCl}
\suppint\left(\bsy{e}^{(n)}_0\right) &=& 
\left[-\left(r^{(n)}_0-1\right)/2,\,\left(r^{(n)}_0+1\right)/2\right],\quad \label{odd_even_0}\\
\suppint\left(\bsy{e}^{(n)}_1\right) &=& 
\left[-r^{(n)}_1/2,\,r^{(n)}_1/2\right]\,. \label{odd_even_1}
\end{IEEEeqnarray}
If $m_n=0$ then applying Lemma~\ref{lem:WS_support_formula} to~(\ref{odd_even_0}) gives
\begin{IEEEeqnarray*}{l}
\suppint\left(\bsy{e}^{(n)}_0\right) \\
\qquad= \left[-\left(r^{(n)}_1+2t^{(n)}-2\right)/2\,,\,\left(r^{(n)}_1+2t^{(n)}\right)/2\right]\,. 
\end{IEEEeqnarray*}
Since $t^{(n)}\geq 1$, comparison with~(\ref{odd_even_1}) yields
\begin{IEEEeqnarray*}{rCl}
-\left(r^{(n)}_1+2t^{(n)}-2\right)/2 & \leq & -r^{(n)}_1/2\quad \mbox{and}\\
\left(r^{(n)}_1+2t^{(n)}\right)/2 & > & r^{(n)}_1/2\,,
\end{IEEEeqnarray*}
which implies $\suppint\left(\bsy{e}^{(n)}_1\right) \varsubsetneq \suppint\left(\bsy{e}^{(n)}_0\right)$.

If $m_n=1$ then applying Lemma~\ref{lem:WS_support_formula} to~(\ref{odd_even_1}) gives
\begin{IEEEeqnarray*}{l}
\suppint\left(\bsy{e}^{(n)}_1\right) \\
\qquad= \left[-\left(r^{(n)}_0+2t^{(n)}-1\right)/2,\,\left(r^{(n)}_0+2t^{(n)}-1\right)/2\right]\,.
\end{IEEEeqnarray*}
Since $t^{(n)}\geq 1$, comparison with~(\ref{odd_even_0}) yields
\begin{eqnarray*}
-\left(r^{(n)}_0+2t^{(n)}-1\right)/2 & < & -\left(r^{(n)}_0-1\right)/2\quad \mbox{and}\\
\left(r^{(n)}_0+2t^{(n)}-1\right)/2 & \geq & \left(r^{(n)}_0+1\right)/2\,,
\end{eqnarray*}
which implies $\suppint\left(\bsy{e}^{(n)}_0\right) \varsubsetneq \suppint\left(\bsy{e}^{(n)}_1\right)$.
\hfill\qed

\section{Half-Sample Symmetric Filter Banks}\label{sec:HS}
We briefly review the definitions from~\cite[Section~IV-B]{Bris09}  of the  group lifting structures for the unimodular HS class, $\mathfrak{H}$.  
In the irreversible case, the scaling matrices are $\mathscr{D} \equiv\left\{\mathbf{D}_K : K\neq 0\right\}$.   The groups of upper (respectively, lower) triangular lifting matrices, 
\mbox{$\mathscr{U} \equiv\upsilon(\mathscr{P})$} (respectively, $\mathscr{L} \equiv\lambda(\mathscr{P})$), are defined by the  same group of whole-sample antisymmetric real Laurent polynomials:
\begin{equation}\label{WA_polys}
\mathscr{P} \equiv \left\{S(z)\in\mathbb{R}[z,z^{-1}] \,:\, S(z^{-1})=-S(z)\right\}\,.
\end{equation}
The set of concentric equal-length base HS filter banks defined in~\cite[Section~IV-B]{Bris09}  is
\begin{equation}\label{B_HS}
\mathfrak{B_H}\equiv \left\{\mathbf{B}\in\mathfrak{H}:\order(B_0)=\order(B_1)\right\}. 
\end{equation}
The lifting cascade group for the irreversible HS class is $\mathscr{C}_{\mathfrak{H}}\equiv\mathscr{\langle U\cup L\rangle}$ and the irreversible HS group lifting structure is $\mathfrak{S_H}\equiv(\mathscr{D},\,\mathscr{U},\,\mathscr{L},\,\mathfrak{B_H})$.

There are no scaling matrices in the reversible case,   $\mathscr{D}_r \equiv\{\mathbf{I}\}$, and the lifting filters~(\ref{WA_polys}) are restricted to those having dyadic rational coefficients.  
The set of base filter banks, $\mathfrak{B}_{\mathfrak{H}_r}$, for the reversible HS class is defined in~\cite[Section~IV-B]{Bris09} to be the set of all matrices in~$\mathfrak{B}_{\mathfrak{H}}$ that {have} dyadic lifting factorizations with no scaling matrices.  
Note that there are no other requirements (e.g., symmetry)   for the dyadic lifting filters that factor reversible, concentric, equal-length HS base filter banks.
The {reversible HS class} is  \emph {defined} to be
\[ \mathfrak{H}_r \equiv \mathscr{C}_{\mathfrak{H}_r}\mathfrak{B}_{\mathfrak{H}_r}
\quad\mbox{where}\quad  \mathscr{C}_{\mathfrak{H}_r}\equiv\langle\mathscr{U}_r\cup \mathscr{L}_r\rangle \,,  \]
and its group lifting structure is $\mathfrak{S}_{\mathfrak{H}_r}\equiv(\mathscr{D}_r,\,\mathscr{U}_r,\,\mathscr{L}_r,\,\mathfrak{B}_{\mathfrak{H}_r})$.

\subsection{Main Uniqueness Result}\label{sec:HS:Main}
Our main result on HS filter banks shows that  irreducible group lifting factorizations in  $\mathfrak{S_H}$ are ``almost'' unique  and that irreducible lifting factorizations in $\mathfrak{S}_{\mathfrak{H}_r}$ are unique.
\begin{thm}[Uniqueness of HS group lifting factorizations]\label{thm:HS_Uniqueness}
Let $\mathfrak{S_H}$ and $\mathfrak{S}_{\mathfrak{H}_r}$ be the group lifting structures defined in~\cite[Section~IV-B]{Bris09}.
Every filter bank in $\mathfrak{H}$ has an irreducible group lifting factorization in $\mathfrak{S_H}$ that is  unique modulo rescaling.
Every filter bank in $\mathfrak{H}_r$ has a unique irreducible group lifting factorization in $\mathfrak{S}_{\mathfrak{H}_r}$.
\end{thm}

\pf 
Existence of irreducible group lifting factorizations for these classes was discussed in~\cite[Section~IV-B]{Bris09}.

Uniqueness:  Let $\mathbf{B}(z)\in\mathfrak{B_H}$.  By the definition of  $\mathfrak{H}$~\cite[Definition~9]{Bris09}, both impulse responses are symmetric about $-1/2$ so~(\ref{B_HS}) implies that $b_0$ and $b_1$ have the same scalar support intervals.  By Lemma~\ref{lem:-.5_poly_suppint} (Section~\ref{sec:HS:Support} below),   $\bsy{b}_0$ and $\bsy{b}_1$ have the same \emph{polyphase} support intervals, verifying~(\ref{equal-order_base}).
Lemma~\ref{lem:HS_support_covering} (Section~\ref{sec:HS:UniqueFact} below) shows that irreducible lifting cascades in  $\mathscr{C}_\mathfrak{H}\mathfrak{B_H}$ satisfy~(\ref{support-covering}), so Lemma~\ref{lem:SufficientConditions} implies that  $\mathfrak{S_H}$ is  order-increasing.
$\mathscr{D}$-invariance  follows from~\cite[Lemma~2]{Bris09}, so~\cite[Theorem~1]{Bris09}  implies that irreducible lifting factorizations in  $\mathfrak{S_H}$ are unique modulo rescaling.

In the reversible case,  $\mathscr{C}_{\mathfrak{H}_r}<\mathscr{C}_\mathfrak{H}$ and $\mathfrak{B}_{\mathfrak{H}_r} \subset \mathfrak{B}_\mathfrak{H}$.  The  \emph{reversible} HS group lifting factorizations thus form a subset of the \emph{irreversible}  factorizations and are therefore unique modulo rescaling.  However, given  two irreducible  factorizations of the same reversible filter bank, 
\[ \mathbf{S}_{N-1}(z) \cdots\mathbf{S}_0(z)\mathbf{B}(z) 
= \mathbf{S}'_{N-1}(z)\cdots \mathbf{S}'_0(z)\mathbf{B}'(z)\,, \]
we have $K'=K=1$ since there are no scaling operations. Thus, $\mathbf{B}'(z)=\mathbf{B}(z)$ and $\mathbf{S}'_i(z) = \mathbf{S}_i(z)$ for all $i$, implying uniqueness of reversible HS group lifting factorizations.
\hfill\qed 

\rem
As in the WS case, Theorem~\ref{thm:HS_Uniqueness} provides a more complete theory of both existence \emph{and} uniqueness for cascade factorizations of HS filter banks than~\cite{NguyenVaidyana:89:Two-channel-perfect-reconstruction-structures}.

\subsubsection{Applications}\label{sec:HS:Main:Applications}
The JPEG~2000 standard does not contain source coding specifications for HS filter banks as a class, in part because of the complications inherent in their lifting factorizations.  Instead, JPEG~2000 Part~2~\cite[Annex~H]{ISO_15444_2} contains specifications for lifting factorizations of \emph{arbitrary} user-defined FIR filter banks, with no assumptions about the symmetries of either the filters or their lifting factors.  Part~2 Annex~H does contain a number of examples of HS filter banks, however, and we can use Theorem~\ref{thm:HS_Uniqueness} to answer some questions about lifting factorization of HS filter banks that arose on the ISO committee during the writing of Part~2.

First let us address the filter bank normalization specifications in JPEG~2000 Part~2.  We mentioned in Section~\ref{sec:Intro:GLS} that~\cite[Theorem~1]{Bris09} yields unique irreducible  factorizations if one normalizes a degree of freedom in the base filter banks, e.g., by fixing a common nonzero value for  $B_0(1)$.
As explained in~\cite{BrisWohl07}, JPEG~2000 Part~2 requires the normalization $H_0(1)=1$,  a similar constraint.  
The JPEG~2000 constraint does \emph{not} imply uniqueness, however.  Let  
\begin{equation}\label{JPEG_FB}
\mathbf{H}(z) = \mathbf{D}_K\,\mathbf{S}_{N-1}(z) \cdots\mathbf{S}_0(z)\, \mathbf{B}(z)
\end{equation}
satisfy $H_0(1)=1$, and pick $K'\neq K$.  Define $\alpha\equiv K/K'$ so that 
$\mathbf{D}_K=\mathbf{D}_{K'}\,\mathbf{D}_{\alpha}$ and substitute in~(\ref{JPEG_FB}).  By~\cite[eqn.~(28)]{Bris09},
\[ \mathbf{H}(z) = \mathbf{D}_{K'}\,\gamma_{\!\alpha}\mathbf{S}_{N-1}(z)\cdots \gamma_{\!\alpha}\mathbf{S}_0(z)\,\mathbf{B}'(z)\,, \]
where  $\mathbf{B}'(z) \equiv \mathbf{D}_{\alpha}\,\mathbf{B}(z)$.
This gain transfer  between the scaling and base matrices leaves $\mathbf{H}(z)$ unchanged, and therefore does \emph{not} violate the  constraint $H_0(1)=1$.  According to Theorem~\ref{thm:HS_Uniqueness}, though, this is the \emph{only} way in which two irreducible HS group lifting factorizations of $\mathbf{H}(z)$ can differ.

As described in~\cite[Section~I-A]{Bris09},  the JPEG~2000 example filter bank in~\cite[Annex~H.4.1.2.1]{ISO_15444_2} contains a lifting factorization of an irreversible 6-tap/10-tap HS wavelet filter bank, $\mathbf{H}_{6,10}(z)$,  that is lifted from a concentric 6-tap/6-tap HS base filter bank, $\mathbf{B}_{6,6}(z)$, by a second-order WA lifting step:
\begin{equation}\label{H610}
\mathbf{H}_{6,10}(z) = \mathbf{S}_0(z)\,\mathbf{B}_{6,6}(z)\,. 
\end{equation}
Interestingly, although $\mathbf{H}_{6,10}(z)$ is irreversible, the scaling factor is trivial ($K=1$), which is why  the scaling matrix has been omitted from~(\ref{H610}).   Factorization~(\ref{H610}) is  unique modulo rescaling by Theorem~\ref{thm:HS_Uniqueness}, which answers in the negative the question of whether $\mathbf{H}_{6,10}(z)$ can be lifted from the Haar filter bank by a clever choice of WA lifting filters.     The same answer holds for the question of whether the 10-tap/18-tap HS filter bank in~\cite[Annex~H.4.1.2.2]{ISO_15444_2}, which is lifted from a 10-tap/10-tap HS base filter bank  by a fourth-order WA lifting filter, can similarly be lifted from the Haar  using WA lifting filters.

Theorem~\ref{thm:HS_Uniqueness} also applies to the group lifting structure $\mathfrak{S}_{\ssst ELASF}$ for the ELASF family  of reversible HS filter banks lifted from the Haar, as defined by M.\ Adams~\cite{Adams02,AdamsWard:03:Symmetric-extension-compatible-reversible-integer-to-integer}.
It was shown in~\cite[Example~6]{Bris09} that the  ELASF filter banks form a subset of the reversible HS class, $\mathfrak{H}_r$, and that every group lifting factorization in $\mathfrak{S}_{\ssst ELASF}$ is a  factorization in $\mathfrak{S}_{\mathfrak{H}_r}$.  Thus,  irreducible group lifting factorizations in $\mathfrak{S}_{\ssst ELASF}$ are unique.

\smallskip

The lemmas needed to verify hypothesis~(\ref{support-covering}) of Lemma~\ref{lem:SufficientConditions} occupy the remainder of Section~\ref{sec:HS}.

\subsection{Polyphase Support Intervals}\label{sec:HS:Support}
Polyphase support intervals for FIR filters centered at $-1/2$ are sufficient to cover HS group lifting structures.  The formulas do not depend on linear phase properties of the filters.

\begin{lem}\label{lem:-.5_poly_suppint}
Suppose $E(z)$ is an even-length FIR filter whose support interval is centered at $-1/2$:
\begin{equation}\label{-.5_suppint}
\suppint(e) = [-r,\,r-1]\quad\mbox{for $r\equiv\supprad(e)\geq 1$.} 
\end{equation}
Then the polyphase support interval of the filter is:
\[ 
\suppint(\bsy{e}) = \left\{
\begin{array}{ll}
[-r/2,\, r/2]  & \mbox{if $r$ is even,} \\
\left[-(r-1)/2,\, (r-1)/2\right]  & \mbox{if $r$ is odd.}
\end{array}
\right.
\]
\end{lem}
\pf There are two cases.
\case{\it  $r$ Even.}\quad
The even polyphase component is 
$e_0(k)=e(2k)$ for \mbox{$-r\leq 2k\leq r-2,$} 
and $e(-r)\neq 0$ while $e(r-2)$ may be zero, so 
\begin{equation}\label{-.5_poly_suppint_even_0}
\suppint(e_0) = [-r/2,\,n]
\end{equation}
for some $n$ with $-r/2\leq n\leq (r-2)/2$.

The odd component is 
$e_1(k)=e(2k-1)$ for \mbox{$-r+1\leq 2k-1\leq r-1$,} 
and \mbox{$e(r-1)\neq 0$} while $e(-r+1)$ may be zero, so
\begin{equation}\label{-.5_poly_suppint_even_1}
\suppint(e_1) = [n,\,r/2]
\end{equation}
for some $n$ with $(-r+2)/2\leq n\leq r/2$.
Combining~(\ref{-.5_poly_suppint_even_0}) and~(\ref{-.5_poly_suppint_even_1}) using~(\ref{poly_supp_int}) yields
\begin{IEEEeqnarray}{rCl}
\suppint(\bsy{e}) &=& \suppint(e_0)\vee\suppint(e_1) \IEEEnonumber\\
&=&   [-r/2,\,r/2]\,.\label{-.5_poly_suppint_even}
\end{IEEEeqnarray}
\case{\it  $r$ Odd.}\quad
The even polyphase component is 
$e_0(k)=e(2k)$ for \mbox{$-r+1\leq 2k\leq r-1$,} 
and $e(r-1)\neq 0$ while $e(-r+1)$ may be zero, so 
\begin{equation}\label{-.5_poly_suppint_odd_0}
\suppint(e_0) = [n,\,(r-1)/2]
\end{equation}
{for some $n$ with $(-r+1)/2\leq n\leq (r-1)/2$.}

The odd component is
$e_1(k)=e(2k-1)$ for \mbox{$-r\leq 2k-1\leq r-2$,} 
and $e(-r)\neq 0$ while $e(r-2)$ may be zero, so 
\begin{equation}\label{-.5_poly_suppint_odd_1}
\suppint(e_1) = [(-r+1)/2,\,n]
\end{equation}
{for some $n$ with $(-r+1)/2\leq n\leq (r-1)/2$.}
Combining~(\ref{-.5_poly_suppint_odd_0}) and~(\ref{-.5_poly_suppint_odd_1}) using~(\ref{poly_supp_int}) yields
\begin{equation}\label{-.5_poly_suppint_odd}
\suppint(\bsy{e}) =   [-(r-1)/2,\,(r-1)/2]\,.
\end{equation}
\hfill\qed

\subsection{The HS Support Interval Formula}\label{sec:HS:Formula}
As in the WS case, the HS support interval formula does not depend on linear phase properties of the filters.
\begin{lem}\label{lem:HS_support_formula}
Let \mbox{$\mathbf{S}_{N-1}(z) \cdots\mathbf{S}_0(z)\,\mathbf{B}(z)\in\mathscr{C}_{\mathfrak{H}}\mathfrak{B_H}$} be irreducible, with intermediate  scalar filters $E^{(n)}_i(z),$ $i=0,\,1.$  Let $r^{(n)}_i$ be the support radius of  $e^{(n)}_i,$ 
and let \mbox{$t^{(n)}\geq 1$} be the support radius of the WA lifting filter $S_n(z)$.
Then $\suppint\left(e^{(n)}_i\right)$ is centered at $-1/2$,
\[ \suppint\left(e^{(n)}_i\right) = \left[-r^{(n)}_i,\,r^{(n)}_i-1\right]\,,\quad i=0,\,1, \]
where
\begin{equation}\label{HS_support_radius}
r^{(n)}_{m_n} = r^{(n)}_{1-m_n} + 2t^{(n)} \quad\mbox{for $n\geq 0$,}
\end{equation}
\begin{equation}\label{HS_support_radius_nonlift}
r^{(n)}_{1-m_n}  = r^{(n-1)}_{m_{n}} + 2t^{(n-1)} \quad\mbox{for $n\geq 1$,}
\end{equation}
with $ r^{(0)}_{1-m_0}=r^{(-1)}$, the common support radius of both base filters, $b_i$.  
\end{lem}
\pf  Induction on $n$.  In all cases, the WA lifting filter is centered at 0:
$\suppint(s_n)=[-t^{(n)},\,t^{(n)}]$.
The concentric equal-length base filter bank, $\mathbf{B}(z)\in\mathfrak{B_H}$, has both filters centered at $-1/2$:
\begin{equation}\label{HS_suppint_bi}
\suppint(b_i) = [-r^{(-1)},\, r^{(-1)}-1]\,,\quad i=0,\,1\,. 
\end{equation}

\case{$n=0.$}\quad
By~(\ref{En_scalar_lifting}) the lifted intermediate filter is 
\begin{equation}\label{HS_lifted_e0}
E^{(0)}_{m_0}(z) = B_{m_0}(z) + S_{0}(z^2)\,B_{1-m_0}(z)
\end{equation}
and its complement is
$E^{(0)}_{1-m_0}(z) = B_{1-m_0}(z)$, implying $r^{(0)}_{1-m_0} = r^{(-1)}$.
By~(\ref{suppint_convolution}) and~(\ref{suppint_interpolation}),
\begin{IEEEeqnarray*}{l}
\suppint\left(S_{0}(z^2)\,B_{1-m_0}(z)\right) \\
\qquad \qquad = [-2t^{(0)},\,2t^{(0)}] + [-r^{(-1)},\, r^{(-1)}-1] \\
\qquad \qquad = [-2t^{(0)}-r^{(-1)},\,2t^{(0)}+r^{(-1)}-1]\,. 
\end{IEEEeqnarray*}
By~(\ref{HS_suppint_bi}), \mbox{$t^{(0)}\geq 1$} implies 
\[ \suppint(b_{m_0}) \subset (-2t^{(0)}-r^{(-1)},\,2t^{(0)}+r^{(-1)}-1)\,, \]
which justifies applying formula~(\ref{suppint_covering}) from Lemma~\ref{lem:suppint} to~(\ref{HS_lifted_e0}):
\begin{eqnarray*}
\suppint\left(e^{(0)}_{m_0}\right) &=& \suppint\left(S_{0}(z^2)\,B_{1-m_0}(z)\right) \\
&=&   [-2t^{(0)}-r^{(-1)},\,2t^{(0)}+r^{(-1)}-1]\,.
\end{eqnarray*}
This shows that $\suppint\left(e^{(0)}_{m_0}\right)$ is centered at $-1/2$ with 
\[ \supprad\left(e^{(0)}_{m_0}\right)=r^{(-1)}+2t^{(0)}\,, \]
verifying~(\ref{HS_support_radius}).
\smallskip

\case{$n>0.$}\quad
The  complement of the lifted filter  is
\[ E^{(n)}_{1-m_n}(z) = E^{(n-1)}_{1-m_n}(z) = E^{(n-1)}_{m_{n-1}}(z) \]
by irreducibility.
Use the induction hypothesis to apply~(\ref{HS_support_radius}) to $E^{(n-1)}_{m_{n-1}}(z)$ and note that $m_n=1-m_{n-1}$:
\begin{equation}\label{rn_induction}
r^{(n)}_{1-m_n} = r^{(n-1)}_{1-m_n} = r^{(n-1)}_{m_{n-1}} = r^{(n-1)}_{m_n} + 2t^{(n-1)} \,,
\end{equation}
which verifies~(\ref{HS_support_radius_nonlift}).  
By~(\ref{En_scalar_lifting}) the lifted intermediate filter is 
\begin{equation}\label{HS_lifted_en}
E^{(n)}_{m_n}(z) = E^{(n-1)}_{m_n}(z) + S_n(z^2)\,E^{(n-1)}_{1-m_n}(z)\,.
\end{equation}
$\suppint\left(e^{(n-1)}_{1-m_n}\right)$ is centered at $-1/2$ by induction, so~(\ref{suppint_convolution}) and~(\ref{suppint_interpolation}) give
\begin{IEEEeqnarray}{l}
\suppint\left(S_n(z^2)\,E^{(n-1)}_{1-m_n}(z)\right) \IEEEnonumber\\
\qquad =  \left[-2t^{(n)},\,2t^{(n)}\right] + \left[-r^{(n-1)}_{1-m_n},\, r^{(n-1)}_{1-m_n}-1\right] \IEEEnonumber\\
\qquad = \left[-2t^{(n)}-r^{(n-1)}_{1-m_n},\, 2t^{(n)}+r^{(n-1)}_{1-m_n}-1\right]\,. \label{HS_suppint_lift}
\end{IEEEeqnarray}
Substitution of~(\ref{rn_induction}) into~(\ref{HS_suppint_lift}) gives
\begin{IEEEeqnarray*}{l}
\suppint\left(S_n(z^2)\,E^{(n-1)}_{1-m_n}(z)\right)  \\
\qquad \qquad = \left[- r^{(n-1)}_{m_n} -  2(t^{(n-1)} + t^{(n)}),\right.\\
\qquad \qquad \qquad \left. r^{(n-1)}_{m_n} + 2(t^{(n-1)} + t^{(n)}) - 1\right]\,.
\end{IEEEeqnarray*}
The induction hypothesis says that $\suppint\left(e^{(n-1)}_{m_n}\right)$ is centered at $-1/2$,
with $t^{(n-1)}\geq 1$ and $t^{(n)}\geq 1$, so
\begin{IEEEeqnarray*}{rCl}
\suppint\left(e^{(n-1)}_{m_n}\right)  &=& \left[- r^{(n-1)}_{m_n},\,  r^{(n-1)}_{m_n}-1\right] \\
&\subset&
\left(- r^{(n-1)}_{m_n} -  2(t^{(n-1)} + t^{(n)})\,,\right.\\
&&\quad\left. r^{(n-1)}_{m_n} + 2(t^{(n-1)} + t^{(n)}) - 1\right)\,.
\end{IEEEeqnarray*}
This justifies applying~(\ref{suppint_covering}) to~(\ref{HS_lifted_en}):
\[ \suppint\left(e^{(n)}_{m_n}\right) = \suppint\left(S_n(z^2)\,E^{(n-1)}_{1-m_n}(z)\right) \,. \]
Formula~(\ref{HS_suppint_lift}) shows that $\suppint\left(e^{(n)}_{m_n}\right)$ is centered at $-1/2$, and substituting
$r^{(n)}_{1-m_n}=r^{(n-1)}_{1-m_n}$ into~(\ref{HS_suppint_lift})  verifies~(\ref{HS_support_radius}):
\[ r^{(n)}_{m_n} = r^{(n)}_{1-m_n}+ 2t^{(n)} \,. \]
\hfill\qed

\subsection{The Support-Covering Property}\label{sec:HS:UniqueFact}
We now prove that $\mathfrak{S_H}$ satisfies the polyphase support-covering hypothesis~(\ref{support-covering}) in Lemma~\ref{lem:SufficientConditions}.
\begin{lem}\label{lem:HS_support_covering}
Let \mbox{$\mathbf{S}_{N-1}(z) \cdots\mathbf{S}_0(z)\,\mathbf{B}(z)\in\mathscr{C}_{\mathfrak{H}}\mathfrak{B_H}$} be irreducible, with intermediate  filters $E^{(n)}_i(z),$ $i=0,\,1.$  Then
\begin{equation}\label{HS_support-covering}
\suppint\left(\bsy{e}^{(n)}_{1-m_n}\right) \varsubsetneq \suppint\left(\bsy{e}^{(n)}_{m_n}\right)
\quad\mbox{for $n\geq 0$.}
\end{equation}
\end{lem}
\pf  Let $r^{(n)}_i$ be the support radius of $e^{(n)}_i$, $i=0,\,1$.  By Lemma~\ref{lem:HS_support_formula} it follows that
\mbox{$r^{(n)}_{m_n} + r^{(n)}_{1-m_n}$} is even, so $r^{(n)}_{0}$ and $r^{(n)}_{1}$ always have \emph{equal} parities.

\case{\it $r^{(n)}_0$, $r^{(n)}_1$ even.}\quad
Regardless of the value of $m_n$,  Lemma~\ref{lem:-.5_poly_suppint} implies
\begin{IEEEeqnarray}{rCl}
\suppint\left(\bsy{e}^{(n)}_{m_n}\right) &=& \left[-r^{(n)}_{m_n}/2,\,r^{(n)}_{m_n}/2\right]\,, \label{even_m_n}\\
\suppint\left(\bsy{e}^{(n)}_{1-m_n}\right) &=& \left[-r^{(n)}_{1-m_n}/2,\,r^{(n)}_{1-m_n}/2\right]\,.\label{even_1-m_n}
\end{IEEEeqnarray}
Applying Lemma~\ref{lem:HS_support_formula} to~(\ref{even_m_n}) gives
\begin{IEEEeqnarray*}{l}
\suppint\left(\bsy{e}^{(n)}_{m_n}\right) \\
\qquad\qquad= \left[-\left(r^{(n)}_{1-m_n}+2t^{(n)}\right)/2,\,\left(r^{(n)}_{1-m_n}+2t^{(n)}\right)/2\right]\,. 
\end{IEEEeqnarray*}
Since $t^{(n)}\geq 1$, comparison with~(\ref{even_1-m_n})  shows that 
\[ \suppint\left(\bsy{e}^{(n)}_{1-m_n}\right) \varsubsetneq \suppint\left(\bsy{e}^{(n)}_{m_n}\right)\,. \]

\case{\it $r^{(n)}_0$, $r^{(n)}_1$ odd.}\quad
Regardless of the value of $m_n$,  Lemma~\ref{lem:-.5_poly_suppint} implies
\begin{IEEEeqnarray}{l}
\suppint\left(\bsy{e}^{(n)}_{m_n}\right) \IEEEnonumber\\
\qquad \; = \left[-\left(r^{(n)}_{m_n}-1\right)/2,\,\left(r^{(n)}_{m_n}-1\right)/2\right]\,, \label{odd_m_n}
\end{IEEEeqnarray}
\begin{IEEEeqnarray}{l}
\suppint\left(\bsy{e}^{(n)}_{1-m_n}\right) \IEEEnonumber\\
\qquad \; = \left[-\left(r^{(n)}_{1-m_n}-1\right)/2,\,\left(r^{(n)}_{1-m_n}-1\right)/2\right] \label{odd_1-m_n}\,.
\end{IEEEeqnarray}
Applying Lemma~\ref{lem:HS_support_formula} to~(\ref{odd_m_n}) gives
\begin{IEEEeqnarray*}{l}
\suppint\left(\bsy{e}^{(n)}_{m_n}\right) \\
\quad= \left[-\left(r^{(n)}_{1-m_n}+2t^{(n)}-1\right)/2,\,\left(r^{(n)}_{1-m_n}+2t^{(n)}-1\right)/2\right]. 
\end{IEEEeqnarray*}
Since $t^{(n)}\geq 1$, comparison with~(\ref{odd_1-m_n})  shows that 
\[ \suppint\left(\bsy{e}^{(n)}_{1-m_n}\right) \varsubsetneq \suppint\left(\bsy{e}^{(n)}_{m_n}\right)\,.\]
\hfill\qed

\section{Conclusions}\label{sec: Conclusions}
The theory of group lifting structures developed in~\cite{Bris09} has been used to prove uniqueness of linear phase lifting factorizations for the two nontrivial classes of linear phase two-channel perfect reconstruction filter banks, the WS and HS classes.  A key technical result,  Lemma~\ref{lem:SufficientConditions}, provides sufficient conditions for inferring that a group lifting structure satisfies the critical polyphase order-increasing property defined in~\cite{Bris09}.
In the WS case, Theorem~\ref{thm:WS_Uniqueness} states that  both reversible and irreversible WS  filter banks have unique irreducible group lifting factorizations using HS lifting filters.  
Theorem~\ref{thm:WS_Uniqueness} implies, e.g., that left matrix factorizations of WS filter banks into HS lifting filters are identical to right matrix factorizations.
The scope of Theorem~\ref{thm:WS_Uniqueness} covers the  specifications for  WS filter banks in Part~2 Annex~G of the ISO/IEC JPEG~2000  standard.

In the HS case, Theorem~\ref{thm:HS_Uniqueness} covers liftings  from equal-length HS base filter banks using WA lifting filters.   In the irreversible case, the WA lifting filters {and} the equal-length HS base  filter banks are unique modulo rescaling.  This  implies, for instance, that  6-tap/10-tap and 10-tap/18-tap HS filter banks specified in JPEG~2000 Part~2 Annex~H \emph{cannot} be lifted from the Haar filter bank using WA lifting filters.  Group lifting factorizations of reversible HS filter banks are  unique,  including M.\ Adams'  ELASF class of reversible HS filter banks lifted from the Haar.  A follow-on paper will characterize the  structure of the groups associated with group lifting structures that satisfy the hypotheses of the uniqueness theorem.

\end{document}